\documentclass[preprint,draft,prl,10pt,twocolumn]{revtex4}

\input{tcilatex}

\begin{document}

\title{Full characterization of a three-photon GHZ state using quantum state
tomography}
\author{K.J. Resch$^{1\ast }$, P. Walther$^{1\ast }$, and A. Zeilinger$%
^{1,2} $}
\affiliation{$^{1}$Institut f\"{u}r Experimentalphysik, Universit\"{a}t Wien,
Boltzmanngasse 5, 1090 Vienna, Austria\\
$^{2}$Institut f\"{u}r Quantenoptik und Quanteninformation, \"{O}%
sterreichische Akademie der Wissenschaften, Austria\\
$^{\ast }$These authors contributed equally to this work.}

\begin{abstract}
We have performed the first experimental tomographic reconstruction of a
three-photon polarization state. \ Quantum state tomography is a powerful
tool for fully describing the density matrix of a quantum system. \ \ We
measured 64 three-photon polarization correlations and used a
``maximum-likelihood'' reconstruction method to reconstruct the GHZ\ state.
\ The entanglement class has been characterized using an entanglement
witness operator and the maximum predicted values for the Mermin inequality
was extracted.
\end{abstract}

\maketitle

As quantum information processing becomes more powerful, more powerful ways
of characterizing quantum states, their properties, and quantum logic
operations acting upon them become necessary. States are characterized using
quantum state tomography; in this technique, projective measurements are
performed on an ensemble of identically prepared quantum states each probing
the state from a different ``perspective''.\ The results of these different
measurements are then used to reconstruct the complete object - the density
matrix of the quantum state. \ Quantum state tomography techniques have
characterized two-photon polarization states with varying amounts of
mixedness and entanglement \cite{whitetomo,kwiatmems}, orbital angular
momentum states \cite{orbital}, Wigner functions of nonclassical states of
light \cite{lvovsky}, trapped ions \cite{blatt}, NMR \cite{chuang}, and
atomic states in optical lattices \cite{atoms}.

It is an experimental challenge to perform tomography on multiphoton states
for two main reasons. Firstly, the number of measurements grows
exponentially with the number of photons; the reconstruction of the
polarization state of $n$ photons necessitates $2^{2n}-1$ $n$-fold
coincidence measurements plus one more for normalization. In addition,
three- and four-photon sources relying on two independent down-conversion
pairs are orders of magnitude dimmer than 2-photon sources. Recently, work
has been reported on multiphoton triggered 2-photon tomography for measuring
subsystems of W-states \cite{harald} and for entanglement purification \cite%
{purification}. In the present work, we describe the first experiment in
which the polarization state of three photons, specifically a
Greenberger-Horne-Zeilinger (GHZ)\ state \cite{GHZpaper,GHZfirst}, has been
completely reconstructed via quantum state tomography.

The polarization state of $3$ photons is described by an $8\times 8$ density
matrix. To reconstruct this matrix requires a set of 64 linearly-independent
3-fold coincidence measurements on a large ensemble of identically-prepared
quantum states. We use the 64 combinations of polarization measurements $%
\left| H\right\rangle $, $\left| V\right\rangle $, $\left| D\right\rangle =1/%
\sqrt{2}\left( \left| H\right\rangle +\left| V\right\rangle \right) $, $%
\left| R\right\rangle =1/\sqrt{2}\left( \left| H\right\rangle -i\left|
V\right\rangle \right) $\ on each photon in the 3-photon coincidence
subspace. Each measurement corresponds to a projection onto a pure state $%
\left| \psi _{\upsilon }\right\rangle $. The number of successful
measurement outcomes is directly proportional to the expectation value $%
\left\langle \psi _{\upsilon }\right| \rho \left| \psi _{\upsilon
}\right\rangle $ with a constant of proportionality given by the state flux, 
$\mathcal{N}$. The density matrix is reconstructed by an inversion
algorithm. Due to experimental imperfections, linear inversion can lead to
unphysical density matrices making calculation of certain important
properties (eg., the concurrence \cite{concurrence}) impossible. These
problems can be avoided using a maximum-likelihood reconstruction \cite%
{maxlike}. We employ the specific method of James et al. \cite{james}.
Uncertainties in quantities extracted from these density matrices were
calculated using a Monte Carlo routine and assumed Poissonian errors.

Our experimental setup is shown in Fig. 1. The second harmonic of a
modelocked Ti:Sapphire laser passes twice through a 2-mm long, type-II
phase-matched $\beta $-barium borate (BBO) crystal and creates
polarization-entangled photon pairs \cite{kwiatdownconv}.\ The transverse
and longitudinal walk-off effects are compensated by additional half-wave
plates (HWP) and extra BBO crystals (COMP). By rotating the polarization
with additional HWP, one for each pair, and tilting the compensation
crystals, any of the four Bell states can be produced. We align the source
such that the Bell states $\left| \phi ^{+}\right\rangle =1/\sqrt{2}\left(
\left| HH\right\rangle +\left| VV\right\rangle \right) $ are created on each
pass of the pump. After the optical elements, including 3-nm bandwidth
filters, the light is coupled into single-mode fibres which direct the light
to fibre-coupled single-photon counting detectors. Experimentally measured
two-fold coincidence rates were $27000s^{-1}$ $(19000s^{-1})$ for the
forward- (backward-) emitted pairs.

When the source creates two pairs of entangled photons in the desired modes,
our initial state is described by $\left| \Psi \right\rangle _{1234}=\left|
\phi ^{+}\right\rangle _{12}\otimes \left| \phi ^{+}\right\rangle _{34},$
where subscripts label the spatial modes. GHZ entanglement is created using
a polarizing beam-splitter (PBS). The PBS\ is a linear optical element that
transmits horizontally-polarized light and reflects vertically-polarized
light. In conjunction with post-selection of those cases where incident
photons in modes $2$ and $3$ emerge from different output ports $A$ and $B$,
the PBS acts as a quantum parity check \cite{paritycheck} -- the photons
must have had the same polarization in the $H/V$-basis. \ When photons in
modes $2$ and $3$ arrive at the PBS simultaneously within their coherence
lengths and we keep only those cases where the photons emerge into different
output ports, our state is conditionally transformed as: $1/2\left( \left|
HH\right\rangle _{12}+\left| VV\right\rangle _{12}\right) \otimes \left(
\left| HH\right\rangle _{34}+\left| VV\right\rangle _{34}\right) \rightarrow
1/\sqrt{2}\left( \left| HHHH\right\rangle _{AB14}+\left| VVVV\right\rangle
_{AB14}\right) $ \cite{GHZmermin}. This four-photon GHZ\ state is reduced to
the three-photon GHZ state $\left| GHZ\right\rangle =1/\sqrt{2}\left( \left|
HHH\right\rangle _{AB1}+\left| VVV\right\rangle _{AB1}\right) $ upon
successful projection of the photon in mode $4$ onto the polarization state $%
\left| D\right\rangle _{4}$.

We perform tomographic measurements using wave plates and polarizers. In
modes $A$ and $B$, quarter-wave plates and freely-rotatable polarizers are
used to make projections onto the polarizations $\left\{ H,V,D,R\right\} $.\
In the actual experiment, modes $1$ and $4$ were folded over one another and
crossed at a second PBS followed by two horizontal polarizers. In this
configuration, only transmitted photons can be detected and this
PBS/polarizer combination acts as a pair of fixed horizontal polarizers. \
Since the polarizer in mode 1 could not be rotated, we required both
quarter- and half-wave plates to make our projections. A four-fold
coincidence between the single-photon counting detectors in modes $A$, $B$, $%
1$, and $4$ signals the following: 1) the creation of two pairs of photons,
2) the multiphoton entangling operation of the PBS and creation of the
four-photon GHZ state 3)\ the projection onto $\left| D\right\rangle _{4}$
to create our three-photon GHZ state and 4) a successful outcome for our
tomographic measurement.

In order to demonstrate that our photons were entangled, we measured modes $%
1 $ and $4$ in the polarization state $\left| D\right\rangle $.\ For the
desired GHZ state, this leaves the photons in modes $A$ and $B$ in the Bell
state $\left| \phi ^{+}\right\rangle _{AB}$. Recall $\left| \phi
^{+}\right\rangle =1/\sqrt{2}\left( \left| HH\right\rangle +\left|
VV\right\rangle \right) =1/\sqrt{2}\left( \left| DD\right\rangle +\left|
AA\right\rangle \right) $, where $\left| A\right\rangle =1/\sqrt{2}\left(
\left| H\right\rangle -\left| V\right\rangle \right) ,$ i.e., its
polarizations are not only perfectly correlated in the $H/V$ basis but also
in the $D/A$ basis. However, if the photons in modes 2 and 3 do not overlap
at the PBS, then the two paths leading to a photon in each of modes $A$ and $%
B$ are distinguishable. This leaves the photons instead in the half-mixed
state $\rho =1/2\left( \left| HH\right\rangle \left\langle HH\right| +\left|
VV\right\rangle \left\langle VV\right| \right) $.\ The correlations of this
state are identical to $\left| \phi ^{+}\right\rangle $ in the $H/V$ basis,
but it has no correlations in the $D/A$ basis. With the polarizers in modes $%
A$ and $B$ set to $\left| A\right\rangle $ and $\left| D\right\rangle $
respectively, we record four-photon coincidences as a function of the delay
mirror position. The results are shown in Fig. 2 and demonstrate the change
in the correlations when the photons from the independent pairs arrive at
the PBS simultaneously. The correlation change is observable as a $\left(
69\pm 5\right) \%$ visibility dip in the four-fold rate. Tomographic
measurements were all taken at the centre of this interference dip where the
interference is a result of the requisite coherent superposition. Note that
a QWP\ was placed in output mode $A$ at 90$^{\circ }$ to compensate
birefringence in the PBS. Each tomographic measurement took 900 seconds and
yielded a maximum 466 four-fold counts for the $\left| VVV\right\rangle $
projection.\ To account for laser fluctuations over this time, we normalized
our data by dividing by the square of the background-corrected singles at
the trigger detector (detector 4). The leading-order background in our
4-fold coincidence signal comes from accidental 2-fold coincidence counts at
the same time as a real 2-folds.\ These rates were estimated from measured
singles and 2-fold coincidence data and subtracted from our signal.

The three-photon density matrix reconstructed from the entire data set of 64
measurements is shown in Fig. 3. On the diagonal of the matrix, the dominant
elements are those corresponding to $\left| HHH\right\rangle $ and $\left|
VVV\right\rangle $. Furthermore, there are strong and mostly positive
coherences between these elements with a small phase shift appearing in the
imaginary part. Thus our state has the qualitative properties of the GHZ\
state.\ The fidelity of our state with the ideal GHZ\ state, $\mathcal{F}%
_{GHZ}=\left\langle GHZ\right| \rho \left| GHZ\right\rangle =(76.8\pm 1.5)\%$
provides quantitative confirmation. Entanglement measures for pure and mixed
2-qubit states are well-known, but the same is not true for 3-qubit states.
With 2 qubits, the states are either separable or entangled, however with 3
qubits there exist two nonequivalent classes of 3-qubit entanglement called
GHZ and W \cite{cirac}. We characterize our state using an entanglement
witness operator which detects GHZ entanglement, $\mathcal{W}=3/4I-P_{GHZ}$,
where $I$ is the identity and $P_{GHZ}$ is a projector onto a GHZ state \cite%
{witnesses,acin}. When the expectation value of this witness operator, Tr$%
\mathcal{W}\rho ,$ is negative, the state is definitely contains GHZ
entanglement. The minimum expectation value for this witness operator and
our density matrix for those states related to $\left| GHZ\right\rangle $ by
local, single-qubit, unitary transformations is $\left\langle \mathcal{W}%
\right\rangle _{MIN}=-0.044\pm 0.016$ which is negative by almost 3-$\sigma $%
.

Fully-reconstructed density matrices are a complete description of the
quantum state and can be used to predict the outcomes of any other
measurements one could have performed. As an explicit example of the
usefulness of such an approach, we will investigate what nonlocal properties
our state could exhibit in a Bell experiment. The classic CHSH-Bell
inequalities are not directly applicable to 3-qubit states \cite{CHSHBell}.
An inequality based on the assumptions of local realism was derived by
Mermin \cite{mermin} and can be stated using the expression $M=\left|
E(ABC^{\prime })+E(AB^{\prime }C)+E(A^{\prime }BC)-E(A^{\prime }B^{\prime
}C^{\prime })\right| ,$ where $A$ $\&$ $A^{\prime }$, $B$ $\&$ $B^{\prime }$%
, and $C$ $\&$ $C^{\prime }$ , can be different polarization measurement
settings made on photons $A$, $B$, and $1$, respectively, and $E$ is the
expectation value of the polarization correlation for those settings \cite%
{cereceda}.\ Any local realistic theory places a strict limit on the maximum
strength of measured correlations such that $M\leq 2$. Our density matrix 
\emph{predicts} the maximum possible Mermin parameter $M_{MAX}=2.73\pm 0.11$%
. Thus our state is able to violate the Mermin inequality and show a
violation of local realism by 6.6 standard deviations.

In this experiment, we have created a three-photon GHZ state, performed a
tomographically complete set of polarization measurements, and reconstructed
its density matrix. The most important consequence of reconstructing the
complete density matrix of a quantum state is that one can use this
reconstruction to find out whether the state fulfills criteria in any
possible experiment. Using an entanglement witness we have shown that our
state lies within the GHZ-class of 3-qubit entangled states. \ Our state is
able to demonstrate a conflict with local realism through a Mermin
inequality. This work is a significant step towards measuring and
understanding real multiparticle entangled states.

We thank A. Steinberg \& M. Mitchell for computer software and \v{C}.
Brukner, C. Ellenor, K. Hornberger, A. Stefanov, A. Steinberg, A. White, \&
M. \.{Z}ukowski for helpful discussions. This work was supported by ARC
Seibersdorf Research GmbH, the Austrian Science Foundation (FWF), project
number SFB 015 P06, NSERC, and the European Commission, contract number
IST-2001-38864 (RAMBOQ).

Figure 1. Experimental Setup. \ An ultra-violet laser pulse makes two passes
through a type-II phase-matched BBO\ crystal. This probabilistically
produces two pairs of photons into the four modes (1-4). A half-wave plate
(HWP) and compensation crystal (COMP) are placed in each path to counter
walk-off effects.\ The polarization of one photon from each pair are rotated
by 90$^{\circ }$\ using HWPs to create a pair of $\left| \phi
^{+}\right\rangle $\ states. These two independent photon pairs are further
entangled using the polarizing beamsplitter (PBS) provided that the photons
in mode $2$\ and $3$\ overlap temporally. The 3-photon GHZ state was
produced between modes $A$, $B$, and $1$\ when photon $4$\ is successfully
projected onto the state $\left| D\right\rangle _{4}$\ signalled by a click
at the trigger detector (TRIG.). Tomographic projections were performed
using quarter-waveplates (QWP) and polarizers (POL.) for photons at
detectors $A$\ $\&$\ $B$\ and a half- or quarter-wave plate in front of a
fixed horizontal polarizer at detector 1. An additional quarter-wave plate
was used in mode $A$\ to cancel an unwanted phase shift from the PBS.

Figure 2.\ Four-fold coincidences versus pump mirror position. \ The source
produces two pairs of photons in the state $\left| \phi ^{+}\right\rangle $\
-- one pair into modes $1$\ $\&$\ $2$\ and the other into modes $3$\ $\&$\ $%
4 $. \ For this measurement, the photons in modes $1$\ $\&$\ $4$\ were
successfully projected onto the state $\left| D\right\rangle $. \ As the
pump mirror is displaced, the overlap between the photons in modes $2$\ $\&$%
\ $3$\ is changed. If the photons do not overlap, then the outputs of PBS
have no correlations in the $D/A$-basis. However, if the photons overlap
perfectly, then the outputs of the PBS are left in the state $\left| \phi
^{+}\right\rangle $\ which has strong correlations. The changes in the
correlations manifest as an interference dip when the output modes $A\ \&\ B$%
\ of the PBS\ are measured in the states $\left| A\right\rangle $\ and $%
\left| D\right\rangle $. \ The visibility of the dip is $\left( 69\pm
5\right) \%$.

Figure 3. Density matrix of a 3-photon GHZ state. \ The real (left-hand) and
imaginary (right-hand) of the density matrix were reconstructed from 64
linearly-independent triggered 3-fold coincidence measurements. \ Large
diagonal elements in the $\left| HHH\right\rangle $\ and $\left|
VVV\right\rangle $\ positions along with large positive coherences indicate
that this state has the qualities of the desired GHZ state. \ Those elements
that have negative values are shown as white-topped bars while those that
are positive are blue-topped.

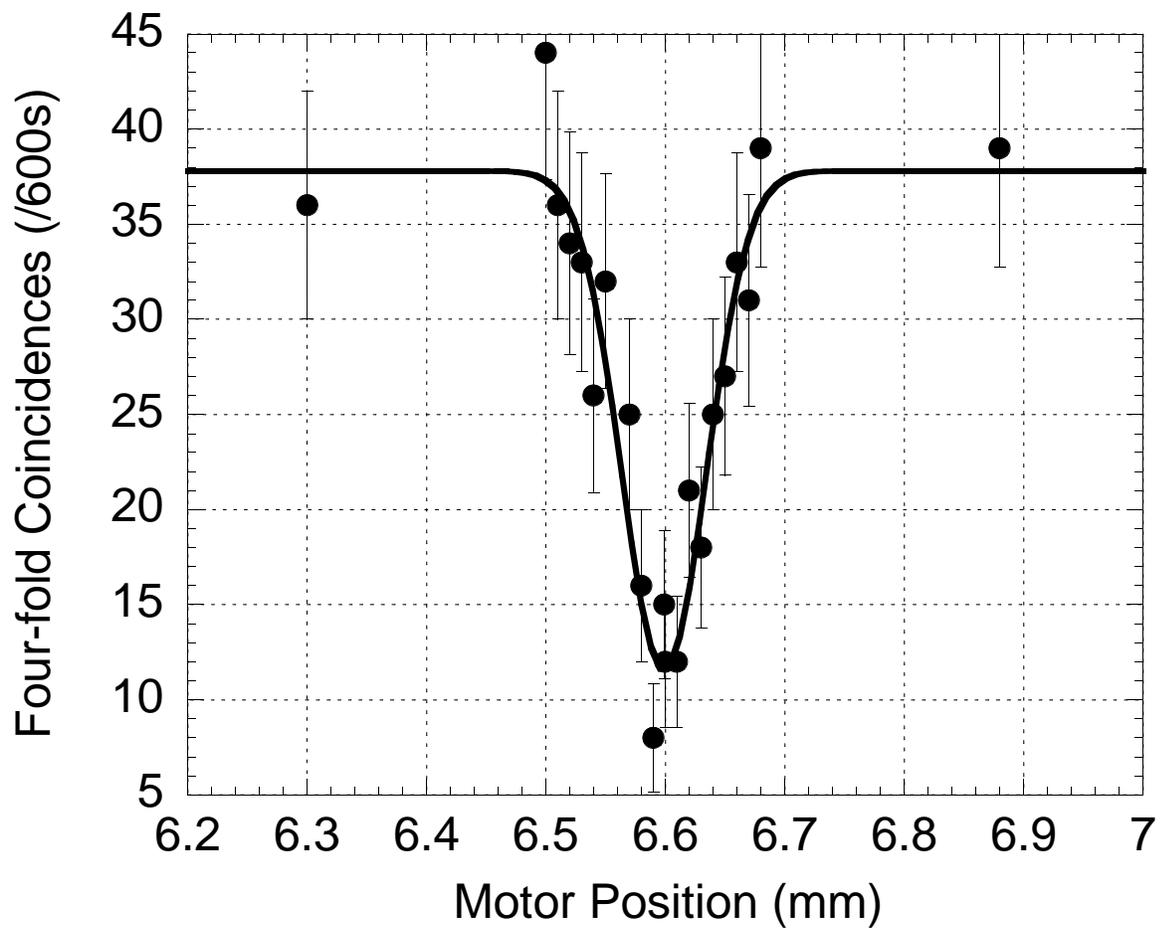
\end{document}